\documentclass[11pt]{article}
\usepackage{amssymb}
\usepackage{epsfig}
\begin{document}
%%%%%%%%%%%%%%
\title{{\bf{\Large Constitutive relations and response parameters in two dimensional hydrodynamics with gauge and gravitational anomalies}}}
\author{  {\bf {\normalsize Rabin Banerjee$^1$}$
$\thanks{E-mail: rabin@bose.res.in}}$~$ 
{\bf and}$~$
{\bf {\normalsize Shirsendu Dey$^1$}$
$\thanks{E-mail: shirsendu12@bose.res.in}},$~$
\\{\normalsize $^1$S. N. Bose National Centre for Basic Sciences,}
\\{\normalsize JD Block, Sector III, Salt Lake, Kolkata-700098, India}
\\\\
}
\maketitle
%%%%%%%%%%%%%%%%%
\begin{abstract}
We obtain the constitutive relations for the stress tensor and gauge current in $(1+1)$ dimensional hydrodynamics in the presence of both gauge and gravitational (conformal as well as diffeomorphism) anomalies. The relations between response parameters and anomaly coefficients are also found. The role of the $Israel~Hartle~Hawking$ vacuum is emphasised. Finally, in the absence of gauge fields, earlier results obtained by a hydrodynamic expansion are reproduced.
\end{abstract}
\section*{•}
   
    The study of hydrodynamics\cite{Landau} in the presence of gauge and gravitational anomalies has recently received considerable attention\cite{Sonsuro:2009}-\cite{Bibhas:2013}. An important aspect of this study is the obtention of constitutive relations that express the stress tensor and gauge current in terms of the fluid variables like fluid velocity, chemical potential and temperature. These relations, in the absence of a gauge field, were obtained earlier by the hydrodynamic expansion approach\cite{Jains:2013,Jensen:2012kj} as well as other approaches\cite{Banerjee:2013qha,Banerjee:psba,Bibhas:2013}. Likewise, connections between the anomaly coefficients
 and certain parameters appearing in the constitutive relations were also found. 
 
 In the presence of gauge fields, however, the above analysis becomes quite non-trivial. Even in $(1+1)$ dimensions, general closed form expressions for the constitutive relations or the connections between the response parameters and the anomaly coefficients have not been presented in the literature.
 
 The present paper precisely addresses this issue. Deviating from the usual gradient expansion technique we exploit the exact form of the $(1+1)$ dimensional effective action that is given in the literature\cite{Polyakov:1981rd,Leut:1985}. This exact result is a consequence of the conformal flatness of the two dimensional metric. From this result the stress tensor and the current are obtained by taking appropriate functional derivatives. It is then possible to express these relations in terms of fluid variables thereby yielding our cherished constitutive relations. These relations involve certain constants which are the integration constants appearing in the solutions of differential equations. They may be fixed by choosing an appropriate boundary condition that will be discussed later on. Finally, we compare our results with the gradient expansion approach. This helps in obtaining the connection between response parameters and anomaly coefficients in the presence of both gauge and gravitational anomalies. 

  \section*{•}
     Consider a $(1+1)$ dimensional static background metric\cite{Banerjee:2013qha}:
  \begin{eqnarray}
  \label{met}
  ds^2=-e^{2\sigma(r)}dt^2 +g_{11}dr^2,
  \end{eqnarray}
  It has a timelike Killing vector and the Killing horizon is given by the solution of the equation $e^{2\sigma}|_{r_0}=0$.
  The U(1) gauge field in $(1+1)$ dimension is given by 
  \begin{eqnarray}
  \label{gf}
  A_a= \left(A_t(r),0\right)
  \end{eqnarray}
  The chemical potential is defined in this way
  \begin{eqnarray}
  \label{chem}
  \mu =A_t(r) e^{-\sigma}.
  \end{eqnarray}
 It is convenient to present the analysis in the null coordinates $(u, v)$ which are defined in terms of (t,r) coordinates as,
  \begin{eqnarray}
  \label{tor}
  u=t-r_*,~~~~v=t+r_*,
  \end{eqnarray}
  where $r_*$ is the tortoise coordinate given by $dr_*=-e^{-\sigma}g_{11}dr$. In this coordinate system the metric takes the following off-diagonal form:
\begin{eqnarray}
\label{uv}
ds^2= -e^{2\sigma}\left(dudv+dvdu\right).
\end{eqnarray}
In order to express the energy momentum tensor and gauge current we introduce the fluid variables. The chemical potential has already been defined in (\ref{chem}). The comoving velocity $u_a$ must satisfy the time-like condition $u^au_a=-1$. Under the background (\ref{uv}) we choose the familiar ansatz\cite{Banerjee:2012iz} that is compatible with this normalisation condition. It is given by,
\begin{eqnarray}
\label{ua}
u_a= -\frac{e^\sigma}{2}(1,1),~~~u^a=e^{-\sigma}(1,1).
\end{eqnarray}
 The velocity dual to $u_a$ is $\tilde{u}_a=\bar{\epsilon}_{ab}u^b$ where $\bar{\epsilon}_{ab}$ is the antisymmetric tensor with $\bar{\epsilon}^{ab}=\frac{\epsilon^{ab}}{\sqrt{-g}}$ and $\bar{\epsilon}_{ab}=\sqrt{-g}\epsilon_{ab}$.Here $\epsilon_{ab}$ is the numerical antisymmetric tensor. In null coordinates these expressions are given by,
 \begin{eqnarray}
 \label{udual}
\epsilon_{uv}=1,~~\epsilon^{uv}=-1,~~~~ \tilde{u}_a=\frac{e^\sigma}{2}(1,-1),~~\tilde{u}^a=e^{-\sigma}(1,-1).
 \end{eqnarray}
 Finally, the fluid temperature T is given in terms of the equilibrium temperature $T_0$ by the Tolman relation\cite{Tolman} $T=T_0 e^{-\sigma}$.

 In order to find the constitutive relations in anomalous hydrodynamics it is first necessary to give the expressions for these anomalies. An anomaly is a breakdown of some classical symmetry upon quantization. Breakdown of diffeomorphism symmetry yields the non-conservation of energy momentum tensor, whereas, trace anomaly is the manifestation of breakdown of conformal invariance upon quantisation. A violation of gauge symmetry is revealed by a non-conservation of the gauge current (gauge anomaly) or, alternatively, by the presence of anomalous terms in the algebra of currents. These anomalous terms are related to the gauge anomaly. 
 The general expressions for the diffeomorphism anomaly, trace anomaly and gauge anomaly, relevant for the present paper, are as follows\cite{Bardeen:1984pm}-\cite{Banerjee:vac},

  \begin{eqnarray}
\label{anomaly1}
{\nabla}_{b}T^{ab}&=&F^{a}_{b} J^{b}+ C_g {\bar{\epsilon}}^{ab}{\nabla}_{b} R,
\end{eqnarray}
\begin{eqnarray}
\label{wi2}
{T}^{a}_{a}&=&C_w R,
\end{eqnarray}
\begin{eqnarray}
\label{wi3}
{\nabla}_{a} J^{a}&=& C_s {\bar{\epsilon}}^{ab}F_{ab}.
 \end{eqnarray}
Here $C_g$, $C_w$, and $C_s$ are the coefficients of the respective anomalies. All these expressions are covariant and are hence termed as the covariant anomalies.
  The first term on the r.h.s of (\ref{anomaly1}) is the usual Lorentz force term whereas the other piece gives the gravitational anomaly in terms of the  Ricci scalar. It may be observed that the structures of the anomalous Ward identities follow from dimensional considerations and covariant transformation properties. No other input is necessary. 
%We consider the general $(1+1)$ dimensional effective action. However we are interested in the covariant form, since energy momentum tensors and currents are only defined modulo local polynomials it is made local by introducing the auxiliary fields $\phi(x)$ and $B(x)$.
 %Variation of the effective action with respect to $g_{ab}$ and $A_a$ leads to the energy momentum tensor$(T_{ab})$ and gauge current$(J_a)$ respectively as done previously in \cite{Banerjee:psba}. The general covariant energy momentum tensor and the gauge current are given by,

 A possible way to obtain the constitutive relations expressing the stress tensor and current in terms of the fluid variables would be to solve the above Ward identities. This is however, an elaborate program. We take advantage of the fact that, due to the conformal flatness of the two dimensional metric,the effective action itself is exactly solvable. Then, by suitable variations of the effective action with respect to the gauge field and the metric, the current and the stress tensor may be determined and eventually recast in terms of the fluid variables. By exploiting our earlier results\cite{Banerjee:psba}, we are able to write the explicit forms for $T_{ab}$ and $J_a$,
\begin{eqnarray}
\label{chi2}
&T_{ab} = \left[C_1T^2-C_w \left(u^c \nabla^d \nabla_d u_c \right)+ \mu^2\left(\frac{1}{2\pi}-C_s\right)\right]g_{ab}
\nonumber
\\
&+\left[2C_w\left(u^c \nabla^d - u^d\nabla^c\right)\nabla_c u_d + 2C_1T^2 + 2\mu^2 \left(\frac{1}{2\pi}-C_s\right)\right]
{u}_a{u}_b
\nonumber
\\
&-\left[2C_g\left(u^c \nabla^d - u^d\nabla^c \right)\nabla_c u_d + C_2T^2+C_s\mu^2\right]\left({u}_a\tilde{u}_b+\tilde{u}_a
{u}_b\right) 
\\
\nonumber
&+ \left\lbrace\left(\frac{C}{\pi}-2(C+P)C_s\right)\frac{T}{T_0}\mu 
+\left(\frac{C^2+P^2}{2\pi}-C_s(C+P)^2\right)\frac{T^2}{{T_0}^2}\right\rbrace \left(2u_au_b+g_{ab}\right)
\\
\nonumber
&+\left\lbrace\left(\frac{P}{\pi}-2(C+P)C_s\right)\frac{T}{T_0}\mu +\left(\frac{CP}{\pi}-C_s(C+P)^2\right)\frac{T^2}{{T_0}^2}\right\rbrace \left({u}_a\tilde{u}_b+\tilde{u}_au_b\right) 
\end{eqnarray}
\begin{eqnarray}
\label{current}
J_a=-2C_s \mu \left(u_a+\tilde{u}_a\right)+\frac{\mu}{\pi}u_a + \left(\frac{C}{\pi}-2(C+P)C_s\right)\frac{T}{T_0} u_a + \left(\frac{P}{\pi}-2(C+P)C_s\right)\frac{T}{T_0} \tilde{u}_a,
\end{eqnarray}
where $C_1$, $C_2$, P and C are arbitrary constants that appear in the solution of the effective action. Incidentally, the nonlocal form of the effective action is converted into a local form by introducing extra auxiliary fields that satisfy certain differential equations. These arbitrary constants are the integration constants related to the solutions of the differential equations\cite{Banerjee:psba}.
 
At this juncture, it is useful to illustrate the compatibility of the constitutive relations(\ref{chi2},\ref{current}) with the Ward identities
(\ref{anomaly1},\ref{wi2},\ref{wi3}). Using,
\begin{eqnarray}
\nabla^a(\mu u_a)=0,~~~~~\nabla^a(\mu \tilde{u}_a)=\frac{-1}{2} \bar{\epsilon}^{ab}F_{ab}.
\end{eqnarray}
  the Ward identity(\ref{wi3}) for the current is easily obtained from(\ref{current}). Similarly, using the relation,
  \begin{eqnarray}
  R=-2 u^a\nabla^b\nabla_au_b.
  \end{eqnarray}
  the trace anomaly(\ref{wi2}) easily follows from(\ref{chi2}).
  Finally, exploiting the identities,
  \begin{eqnarray}
  \nabla^a \mu + \mu u^b\nabla_b u^a&=& F^{ab} u_b
  \\
  \nonumber
\nabla^a\left[e^{-2\sigma}\left(2u_au_b+g_{ab}\right)\right]&=&\nabla^a\left[
 e^{-2\sigma} \left({u}_a\tilde{u}_b+\tilde{u}_au_b\right)\right]=0.
  \end{eqnarray}
 and after some algebra, the Ward identity (\ref{anomaly1}) is reproduced.

 We now choose a boundary condition to fix the arbitrary constants. It may be recalled that, in the absence of a gauge field, the $Israel~Hawking~Hartle$ type of boundary condition\cite{Bibhas:2013} reproduced the results obtained by the hydrodynamic expansion\cite{Jensen:2012kj}. This vacuum required that the stress tensor or the current in Kruskal coordinates corresponding to both the outgoing and ingoing modes must be regular near the horizon. It is thus essential to choose $J_u\rightarrow 0$, $J_v\rightarrow 0$, $T_{uu}\rightarrow 0$ and $T_{vv}\rightarrow 0$ near the horizon. To implement these features, the metric (\ref{met}) is considered to be a solution of the Einstein equation. Also, since it is static, event and Killing horizons will coincide\cite{carter} to give the condition $\frac{1}{g_{11}}= e^{2\sigma}|_{r_0}=0$, where $r=r_0$ is the location of the horizon. The constants C and P pertaining to the gauge sector are explicitly determined by enforcing $J_u|_{r_0}=J_v|_{r_0}\rightarrow 0$ to yield,
 \begin{eqnarray}
 \label{jujv}
 P-C=A_t(r_0)=\mu e^{\sigma}|_{r_0}=0,~~ for~~J_u\rightarrow0
 \\
 \nonumber
 P+C=-A_t(r_0)=-\mu e^{\sigma}|_{r_0}=0,~~ for~~ J_v\rightarrow0.
 \end{eqnarray}

The trivial solution is $P=C=0$. Once C and P have been fixed, the constants $C_1$ and $C_2$ relevant for the gravitational sector may be similarly obtained. The result is\cite{Bibhas:2013},
\begin{eqnarray}
\label{c1c2}
C_1=4\pi^2C_w,~~~~C_2=8\pi^2C_g.
\end{eqnarray}
The energy momentum tensor(\ref{chi2}) and gauge current(\ref{current}) after enforcing the constants (\ref{jujv},\ref{c1c2}) are expressed as,\begin{eqnarray}
\label{tab}
&T_{ab} = \left[4\pi^2C_wT^2-C_w \left(u^c \nabla^d \nabla_d u_c \right)+ \mu^2\left(\frac{1}{2\pi}-C_s\right)\right]g_{ab}
\nonumber
\\
&+\left[2C_w\left(u^c \nabla^d - u^d\nabla^c\right)\nabla_c u_d + 8\pi^2C_wT^2 + 2\mu^2 \left(\frac{1}{2\pi}-C_s\right)\right]
{u}_a{u}_b
\nonumber
\\
&-\left[2C_g\left(u^c \nabla^d - u^d\nabla^c \right)\nabla_c u_d + 8\pi^2C_gT^2+C_s\mu^2\right]\left({u}_a\tilde{u}_b+\tilde{u}_a
{u}_b\right) 
\end{eqnarray}
\begin{eqnarray}
\label{ja}
&J_a=-2C_s\mu \left(u_a+\tilde{u}_a\right)+\frac{\mu}{\pi}u_a
\end{eqnarray} 
These constitutive relations are new findings. In the absence of the gauge fields there exists only the first relation (\ref{tab}) with $\mu=0$. It correctly reproduces earlier findings in the literature\cite{Jains:2013,Jensen:2012kj,Bibhas:2013}.

It is now possible to compare our results with the gradient expansion approach\cite{Jensen:2012kj}. This will also immediately fix the response parameters. It is pertinent to point out that one of these parameters in the presence of the U(1) gauge field could not be fixed by the gradient expansion approach. This was one of the reasons that constitutive relations could not be completely determined in that approach. Since these relations have now been obtained in (\ref{tab},\ref{ja}), it is possible, by a comparison, to fix the response parameters.

In the derivative expansion approach, the covariant gauge current is expressed as\cite{Jensen:2012kj},
 \begin{eqnarray}
 \label{jajnsn}
 J_a = -2C_s\mu \tilde{u}_a+{\left(\frac{{\partial}P}{{\partial}\mu}- \frac{{a_2}{'}}{T^2}S_2+\frac{4a_2}{T}S_4\right)} u_a
 \end{eqnarray}
 where,
 \begin{eqnarray}
 \label{p}
 P=T^2p_0(\frac{\mu}{T})
 \end{eqnarray}
 and $S_2$, $S_4$ are some combinations of the gauge field that occur in the second order expansion. The coefficient $a_2$ (as well as its derivative ${a_2}{'}$) and the response parameter $p_0$ are undetermined functions of ${(\frac{\mu}{T})}$.

 Comparing(\ref{jajnsn},\ref{p}) with(\ref{ja}), it is found that,
 \begin{eqnarray}
 \label{comj}
 &&\frac{\partial P}{\partial\mu}=T^2\frac{\partial p_0}{\partial\mu}=\left(-2C_s+\frac{1}{\pi} \right)\mu
\nonumber
\\
&&a_2={a_2}{'}=0.
\end{eqnarray}
leading to the solution,
\begin{eqnarray}
\label{po2}
p_0=\left(\frac{1}{2\pi}-C_s\right)\frac{\mu^2}{T^2}+ Q
\end{eqnarray}
where Q is an integration constant which is determined subsequently by comparing expressions for $T_{ab}$ in the gradient expansion approach, as given in\cite{Jensen:2012kj}, subjected to the relations(\ref{comj}), with (\ref{tab}). The result in the gradient expansion approach simplifies to,
\begin{eqnarray}
\label{tab2}
&T_{ab} = \left(p_0T^2-C_w u^c \nabla^d \nabla_d u_c \right)g_{ab}
+\left[2C_w\left(u^c \nabla^d - u^d\nabla^c\right)\nabla_c u_d + 2p_0T^2\right]
{u}_a{u}_b
\\
\nonumber
&-\left[2C_g\left(u^c \nabla^d - u^d\nabla^c \right)\nabla_c u_d-\bar{C_{2d}}T^2+C_s{\mu}^2\right]\left({u}_a\tilde{u}_b+\tilde{u}_a
{u}_b\right) 
\end{eqnarray}

Now comparing (\ref{po2}) and (\ref{tab2}) with (\ref{tab}) immediately yields,
\begin{eqnarray}
\label{p0c2d}
p_0=4\pi^2C_w+(\frac{1}{2\pi}-C_s)\frac{\mu^2}{T^2}
\end{eqnarray}
\begin{eqnarray}
\label{p0c2d1}
\bar{C}_{2d}=-8\pi^2C_g
\end{eqnarray}
 The relation (\ref{p0c2d}) is a new finding. In the absence of gauge field($\mu=0$), it reproduces earlier results\cite{Banerjee:psba}. Also, as claimed in\cite{Jensen:2012kj}, the relation (\ref{p0c2d1}) does not incur any correction in the presence of the  gauge field.

Let us summarise our findings. We have constructed the constitutive relations for the stress tensor and gauge current in $(1+1)$ dimensional hydrodynamics in the presence of  gauge, conformal and gravitational anomalies. Also, we were able to provide relations connecting the anomaly coefficients with certain response parameters. Both these results are new findings. As a consistency check we reproduced the known expressions in the absence of the gauge field.

A standard approach in the context of anomalous hydrodynamics is the derivative expansion method. While such an approach seems mandatory in higher (greater than $(1+1)$) dimensions, the same is not true for the $(1+1)$ dimensional example. This is due to the conformal flatness of the metric which leads to an exact expression for the effective action. From this expression both the stress tensor and the gauge current may be exactly evaluated by taking appropriate functional derivatives. We take recourse to such an approach, more so because in the presence of gauge fields the hydrodynamic expansion is laced with great difficulties.

   By exploiting our previous results\cite{Banerjee:2013qha, Banerjee:psba} we succeeded in obtaining the cherished constitutive relations. The compatibility of these relations with the anomalous Ward identities was explicitly demonstrated.
   
   The constitutive relations involved several constants that were an outcome of the solutions of differential equations. By choosing the $Israel~Hartle ~Hawking$ boundary condition, all these constants were determined. The efficacy of this boundary condition was earlier discussed\cite{Bibhas:2013} in the absence of gauge fields and led to results that were identical with the hydrodynamic expansion technique. Incidentally, as shown in \cite{Bibhas:2013}, the method of imposing the boundary condition was similar to the derivation of the Cardy formula. It was reassuring to note that the same boundary condition provided consistent results in the presence of both gauge and gravitational fields. This consistency was linked to the fact that our results could be satisfactorily matched with those found by the hydrodynamic expansion by providing additional inputs, namely, the connection of response parameters with anomaly coefficients(\ref{p0c2d},\ref{p0c2d1}) and the identification of certain variables (\ref{comj}). Indeed, it was because of this lack of information that the hydrodynamic expansion was unable to yield constitutive relations in the presence of the gauge anomalies.
   
   As a final remark we note that the choice of the vacuum appears to play a significant role in anomalous fluid dynamics.To what extent this role will exist in higher dimensions is a question for the future.
   
   \section*{Acknowledgement}
   One of the authors S.D would like to thank Dr. Bibhas Ranjan Majhi for some useful discussion.

\end{document}